\documentclass[aps,pra,showpacs,twocolumn]{revtex4-1}
\usepackage{graphicx}
\usepackage[usenames]{color}
\usepackage{amssymb,amsmath}

\newcommand{\eq}[1]{Eq.~(\ref{#1})}
\newcommand{\fig}[1]{Fig.~\ref{#1}}
\newcommand{\be}[1]{\begin{equation}\label{#1}}
\newcommand{\ee}{\end{equation}}

\begin{document}

\title{ A toolkit for semiclassical computations for strongly-driven molecules:  ``frustrated" ionization  of H$_{2}$  driven by elliptical laser fields}

\author{H. Price, C. Lazarou, and A. Emmanouilidou\email{}}

\affiliation{
Department of Physics and Astronomy, University College London, Gower Street, London WC1E 6BT, United Kingdom\\
}

\begin{abstract}
We study the formation of highly excited neutral atoms during the break-up of strongly-driven molecules. Past work on this significant phenomenon has shown that during the formation of highly excited neutral atoms ($\mathrm{H^{*}}$) during the break-up of H$_{2}$ in  a linear laser field the electron that escapes does so either very quickly or after remaining bound for a few periods of the laser field. Here,
we address the  electron-nuclear dynamics in $\mathrm{H^{*}}$ formation  in elliptical laser fields, through  Coulomb explosion. We show that with increasing ellipticity  two-electron effects are effectively ``switched-off".
We perform these studies using a toolkit we have developed for semiclassical computations for strongly-driven multi-center molecules. This toolkit includes the formulation of the probabilities of strong-field phenomena  in a transparent way. This allows us to  identify the shortcomings of currently used initial phase space distributions for the electronic degrees of freedom. In addition, it includes a 3-dimensional method for time-propagation that fully accounts for the Coulomb singularity.
This technique has been previously developed in the context of celestial mechanics and we currently adopt it  to strongly-driven systems.  Moreover, we allow for tunneling during the time-propagation. We find that this is necessary in order to accurately
describe the fragmentation of strongly-driven  molecules.

  \end{abstract}
\date{\today}

\maketitle
\section{Introduction}
A wealth of physical phenomena take place during the fragmentation of strongly-driven molecules by intense infrared laser fields.
 Such phenomena include bond-softening and above-threshold dissociation \cite{Guisti-Suzor1990PRL,ZavriyevPRA1990}, molecular non-sequential double ionization (NSDI) \cite{Staudte2002,Sakai2003PRA,Alnaser,Niikura} and enhanced ionization (EI) \cite{bandrauk1996,Niikura,Wu2012}.
Exploring the interplay of electronic and nuclear motion during the break-up of strongly-driven molecules is a task of great interest. Understanding break-up dynamics  paves the way
for controlling and imaging molecular processes  \cite{Imaging}; it is, however,
a highly challenging task due to the many degrees of freedom involved.

The formation of highly excited neutral fragments  in linearly polarized laser fields 
  has attracted a lot of interest in the last few years \cite{nubbe, JMcKenna, Doerner2011,Eichmann1,Eichmann2}.  In \cite{Emmanouilidou2012} we reported
a theoretical study of the mechanisms of this ``frustrated"---since only one electron eventually escapes---double ionization  process. The  break-up of $\mathrm{H_{2}}$  into a proton, a Rydberg atom ($\mathrm{H^{*}}$) and an escaping electron through Coulomb explosion of the nuclei is a significant phenomenon. It accounts roughly for 10\% of all possible events during the  break-up of $\mathrm{H_{2}}$. Thus, to obtain a complete picture of the break-up of   $\mathrm{H_{2}}$ it is important to also understand the dynamics leading to $\mathrm{H^{*}}$ formation. For linear fields, we have shown that $\mathrm{H^{*}}$ formation takes place through two distinctly different routes  depending on which one of the two ionization steps is ``frustrated".

Currently, quantum mechanical computations in 3-dimensions for $\mathrm{H^{*}}$ formation during the break-up of strongly-driven $\mathrm{H_{2}}$ are out of reach.
In this work, we present a toolkit for 3-dimensional (3-d) semiclassical calculations for the break-up of strongly-driven multi-center molecules. Previous semiclassical 3-d models did not account for nuclear motion; they used fixed-centers to elucidate double ionization in strongly-driven diatomic molecules \cite{Chinesemolecule2,Chinesemolecule1,Emmanouilidou2009}. The important aspects of the toolkit we present are the following: we formulate the computation of probabilities of strong-field phenomena in a transparent way. This allows us to  identify the shortcomings of currently used initial phase space distributions for the electronic degrees of freedom; these shortcomings are more evident when transitioning from the tunneling to the over-the-barrier intensity regime. Moreover, we use a  3-d method for time-propagation that explicitly accounts for the Coulomb singularity while treating two-electron effects
as well as nuclear and electronic motion at the same time. This 3-d method involves the global regularization scheme described in \cite{Heggie1974} as well as a time-transformed leapfrog propagation technique \cite{Mikkola2002} in conjunction with the Bulirsch-Stoer method \cite{Stoer,Press2007}. This technique has been developed in the context of gravitational few-body systems \cite{Mikkola1999, Mikkola2002, Mikkola2013} and we currently adopt it to treat
strongly-driven molecules. The advantage of this latter  propagation technique over the one we previously used  in \cite{Emmanouilidou2009,Emmanouilidou2012} is that it is numerically more robust with a smaller propagation error. The reason is that in the current technique the masses do not enter in the time-transformation resulting in a more accurate treatment of many-body systems with large mass ratios \cite{Mikkola2002}. Another important element of this toolkit is allowing for tunneling during the propagation, that is, the time-propagation is not fully classical. We find this to be necessary in order to accurately
describe phenomena related to enhanced ionization during the  fragmentation of strongly-driven molecules.

Elucidating   the electron dynamics and its interplay with nuclear motion in $\mathrm{H^{*}}$ formation during the break-up of $\mathrm{H_{2}}$ by elliptical laser fields is a challenging problem.
We do so for two intensities: one  intensity in the tunneling and one in the over-the-barrier regime.  We show how the degree of ellipticity of the laser field changes the contribution of each of the  pathways leading to $\mathrm{H^{*}}$ formation. Specifically, we
show that by using an  elliptical field we can ``switch-off" the contribution of the pathway where two-electron effects are important (pathway B). We find that one-electron effects (pathway A) prevail with increasing ellipticity. Moreover, we discuss how the observable 2-d momentum distribution of the escaping electron in $\mathrm{H^{*}}$ formation changes with increasing ellipticity. Finally,
we identify  the tunneling site of the initially bound electron (electron 2).

\section{The model}

We consider an elliptically polarized laser field with its $\hat{z}$ axis parallel to the molecular axis.
We consider a laser field  $\vec{E}(t)=E_{0}(t)(\cos({\omega t})\hat{z}+\epsilon\sin({\omega t}) \hat{x})$ at 800 nm  corresponding to
$\omega= 0.057$ a.u. (a.u. - atomic units), with  $\epsilon$ the ellipticity of the laser field.
 In the current work, we consider a pulse envelope $E_{0}(t)$ of the form $E_{0}(t) = E_{0}$
for $0 < t < 10T$ and $E_{0}(t) = E_{0}\cos^2(\omega(t-10T )/8)$ for
$10T < t < 12T$, with T the period of the field. In what follows for all our calculations the laser field intensities considered refer to $E_{0}^2. $We start the propagation at $\omega t_0 = \phi_0$, where the initial phase of the laser field $\phi_0$ is chosen in the interval $[-\pi/2, 3\pi/2]$.
$\phi_{0}$ can be selected randomly. For computational efficiency, in the current work, we select  equally spaced $\phi_{0}$.
For each  $\phi_{0}$ we set up the initial phase space distribution and compute
the probability for the process under consideration $P^{proc}_{\phi_{0}}$; in the current work this process is the formation of highly excited neutral fragments.
We then compute the total probability for the process of interest by averaging over all $\phi_{0}$ as follows:
\begin{equation}
P^{proc}=\frac{\sum_{\phi_{0}} P^{proc}_{\phi_{0}}\times \Gamma(\phi_0)}{\sum_{\phi_{0}} \Gamma(\phi_0)},
 \label{eq:1}
 \end{equation}
 where $\Gamma(\phi_{0})$ is the ionization rate for field strength $\vert\bar{E}(t_0)\vert$, see Appendix \ref{ionrate}.
We are justified in computing $P^{proc}_{\phi_{0}} $ for each $\phi_{0}$ separately, since, for any process under consideration,  the probabilities at different  $\phi_{0}$ are independent of each other. Note that  computing the total probability using \eq{eq:1}  is different than the method presented in \cite{Chinesemolecule1}, which is equivalent to:
\begin{equation}
P^{proc}=\frac{\sum_{\phi_{0}^{proc}} \Gamma(\phi_{0}^{proc})}{\sum_{\phi_{0}} \Gamma(\phi_0)}
 \label{eq:2}
\end{equation}
In \eq{eq:2} each initial condition is created at a different randomly selected $\phi_{0}$ in $[-\pi/2, 3\pi/2]$;  $\phi_{0}^{proc}$ denotes the $\phi_{0}$ of a trajectory labeled as $\it{proc}$, for instance,  a ``frustrated" ionization trajectory.  For an intensity in the over-the-barrier regime, care must be taken when using  \eq{eq:2}
 to correctly account for the different normalization constants of the electronic initial phase space distributions in the below- and the over-the-barrier intensity regime. Note that this is not an issue when using \eq{eq:1}, since  $P^{proc}_{\phi_{0}} $
is computed for each $\phi_{0}$ separately. We have checked that both \eq{eq:1} and \eq{eq:2} give the same results, when the different normalization constants are properly accounted for in \eq{eq:2}.

\subsection{Initial phase space distribution of the electrons}

 We next discuss how to set up the initial phase space distribution for the below- and
the over-the-barrier intensity regimes for the electronic degrees of freedom.
 If the instantaneous field strength at $\phi_0$ is smaller than the threshold field strength for over-the-barrier ionization, we assume that one electron (electron 1) tunnel ionizes, i.e., tunnels through the field-lowered Coulomb potential to the continuum with an initial velocity distribution that is perpendicular to the direction of the  field  \cite{Chineseelliptical}. It is interesting to note that this first assumption has been very recently verified experimentally for the case of strongly-driven Ar \cite{tunnelexperiment}. We take the electron's initial position to be the classical exit point, see Section \ref{exitpoint}. To describe the initially bound electron (electron 2), we use a one-electron microcanonical distribution \cite{Olson}. For the ionization rate, $\Gamma({\phi_{0}})$, we use the semiclassical formula derived in \cite{Murray2011}, see Appendix \ref{ionrate}.

If the instantaneous field strength at $\phi_0$ corresponds to the over-the-barrier intensity regime, then we employ two different methods to set-up the initial phase space distribution for the electrons. In method 1, a double microcanonical distribution is used  \cite{Olson} for the two electrons, which has already been used  in \cite{Chinesemolecule2,Chinesemolecule1} and in previous work of ours  \cite{Emmanouilidou2009,Emmanouilidou2012}. An alternative method (method 2) is described in what follows:
for electron 1 we assume that it  tunnel ionizes at  the maximum of the field-lowered Coulomb potential; its kinetic energy is equal to the difference between the first ionization energy and the maximum of the field-lowered Coulomb potential, for details see \ref{exitpointOVB}. For electron 2 we employ the same one-electron microcanonical distribution as  for the below-the-barrier intensity regime.

\begin{figure}[h]
\centerline{\includegraphics[scale=0.2,clip=true]{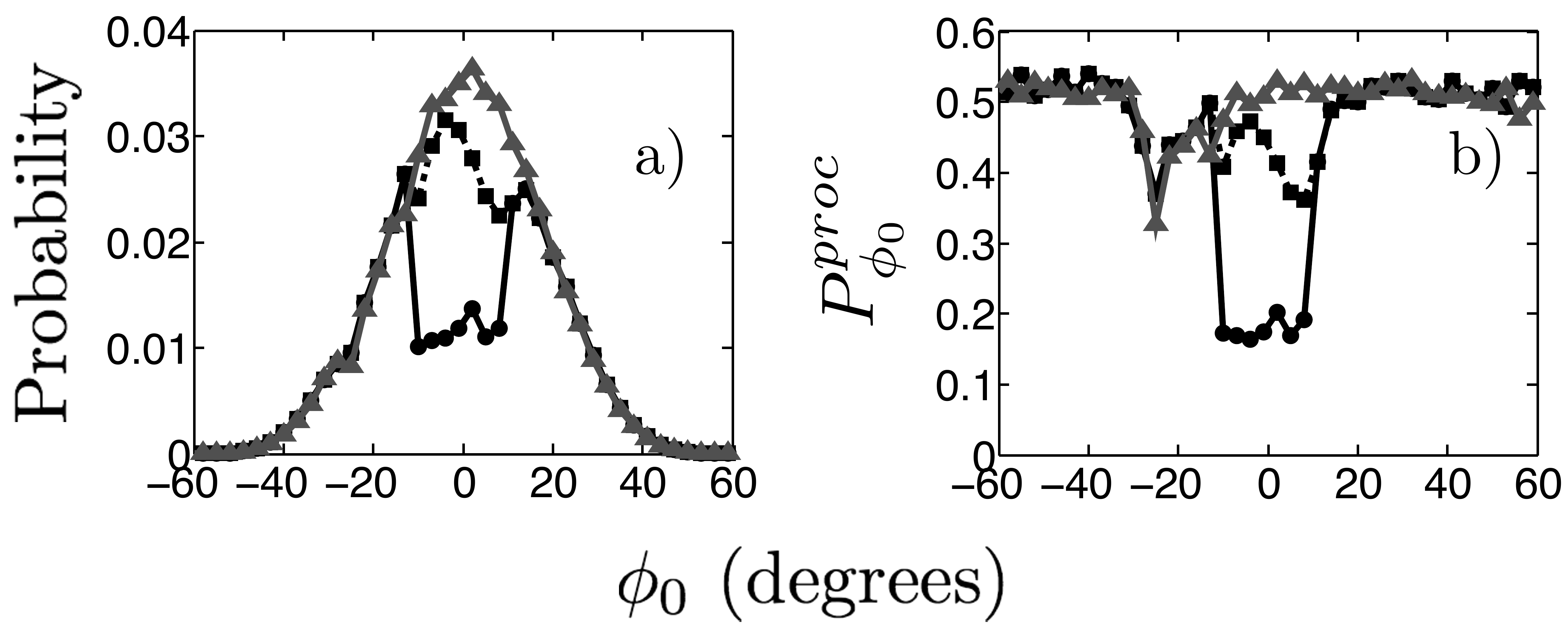}}
\caption{ Double ionization events of strongly-driven H$_{2}$  for an intensity in the below-the-barrier intensity regime, $2.03\times10^{14}$ W/cm$^2$, (grey solid line with full triangles), and for an intensity in the over-the-barrier intensity regime, $2.14\times10^{14}$ W/cm$^2$, using method 1 (black solid line with full circles) and method 2 (black dashed line with full squares): the distribution for the initial phase $\phi_0$ a) and $P^{proc}_{\phi_{0}}$ as a function of $\phi_{0}$ b). }
\label{fig:phasealldouble}
\end{figure}

We use both methods and compare the results for the probabilities for double ionization and frustrated ionization
for an intensity just below ($2.03\times10^{14}$ W/cm$^2$) and just above ($2.14\times10^{14}$ W/cm$^2$) the threshold intensity for over-the-barrier ionization. One expects that, for each process, the probabilities at these two similar intensities should be very close to each other. In   table \ref{tab:prob}, we show that this condition  is satisfied best when using  method 2. Moreover, for the above two intensities, in Fig. \ref{fig:phasealldouble} a) we plot the distribution of the initial phase of the laser field $\phi_0$, that is we plot $P^{proc}_{\phi_{0}}\times \Gamma(\phi_0)/\sum_{\phi_{0}} \Gamma(\phi_0)$ for  double  ionization events.  We find that the distributions for  $2.03\times10^{14}$ W/cm$^2$ and $2.14\times10^{14}$ W/cm$^2$ are more similar  when using method 2 rather than method 1 for the over-the-barrier intensity regime.   That method 2 is better than method 1 can also be seen in Fig. \ref{fig:phasealldouble} b) by plotting the probability $P^{proc}_{\phi_{0}}$ as a function of $\phi_{0}$. It can be clearly seen that when using method 1 $P^{proc}_{\phi_{0}}$  drops sharply in magnitude for $\phi_{0}$ corresponding to field strengths in the over-the-barrier-intensity regime. We, therefore, adapt method 2 in our calculations for the over-the-barrier intensity regime.

\begin{table}
\centering
\begin{tabular}{  | l | c  | c | c | }
  \hline
  I (W/cm$^2$) & $2.03\times10^{14}$ & $2.14\times10^{14}$ & $2.14\times10^{14}$ \\
  & & (Method 1) & (Method 2) \\
 \hline
Double ion. &49\%&34\%&45\% \\
  \hline
 Frustrated ion. &6.3\%&4.7\%&5.6\% \\
  \hline
  \end{tabular}
 \caption{The total probabilities for double ionization and ``frustrated" ionization of strongly-driven H$_{2}$ for an intensity     in the below-the-barrier regime, $2.03\times10^{14}$ W/cm$^2$, and for an intensity in the over-the-barrier regime, $2.14\times10^{14}$ W/cm$^2$. For this latter intensity the probabilities were obtained using   method 1 (third column), and method 2 (fourth column).}
 \label{tab:prob}
\end{table}

\subsubsection{Exit point of tunneling electron for the below-the-barrier intensity regime}\label{exitpoint}
Assuming electron 1 tunnel ionizes with zero momentum along the field direction, we compute the position where electron 1 exits from the field-lowered Coulomb potential using the following equation:

\begin{eqnarray} \label{eq:App3}
V({r}_{1,\parallel},t)&=&-\frac{Z_1}{\left\vert\bar{r}_{1}-\bar{R}_{1}\right \vert}-\frac{Z_2}{\left \vert\bar{r}_{1}-\bar{R}_2\right \vert}+\int\frac{\vert\Psi(\bar{r}_2)\vert^2}{\vert\bar{r}_1-\bar{r}_2\vert}\textrm{d}\bar{r}_2\nonumber \\ \\
&&+\bar{r}_1\cdot\bar{E}(t)=-I_{p1}\nonumber
\end{eqnarray}
computed at $t=t_{0}$. We solve \eq{eq:App3} for ${r}_{1,\parallel}$, the component of $\bar{r}_{1}$ along the direction of the field, while setting equal to zero the component of $\bar{r}_{1}$ perpendicular to the field;  $I_{p1}$ is the first ionization potential, which for  $\mathrm{H_2}$ is equal to 0.57 a.u.
The integral in Eq.~(\ref{eq:App3}) accounts for the screening effect from the bound electron 2.
Expressing the wave function $\Psi(\bar{r}_2)$ of the bound electron 2 in terms of Gaussians
we obtain an analytic expression for this integral.  $\Psi(\bar{r}_2)$ is the  1$\sigma_g$ wave function of $H_{2}^{+}$ at the equilibrium distance of $\mathrm{H_2}$, which we obtain using  MOLPRO---a quantum chemistry package \cite{MOLPRO_brief}. To obtain a relatively simple analytic expression for the integral in \eq{eq:App3} we expand the wave function in terms of s-symmetry Gaussian functions:
	\begin{equation}\label{eq:App5}
		\Psi(\bar{r}_2)=\sum_{j}\sum_n c_{j,n}\phi_{j,n}(\bar{r}_2-\bar{R}_j),
	\end{equation}
with $\phi_{j,n}(\bar{r})$ the contracted s-type functions
	\begin{equation}\label{eq:App6}
		\phi_{j,n}(\bar{r})=\sum_i d_{j,n,i}\Bigg(\frac{2\alpha_{j,n,i}}{\pi}\Bigg)^{3/4}e^{-\alpha_{j,n,i}\bar{r}^2},
	\end{equation}
and $\bar{R}_j$ the position vectors of the nuclei.  Expanding in s-symmetry Gaussian functions is a very good approximation for the wave function currently under consideration. The final expression for the screening potential due to electron 2 is
	\begin{eqnarray} \label{eq:App7}
		 \int\frac{\vert\Psi({\bar r}_2)\vert^2}{\vert{\bar r}_1-{\bar r}_2\vert}\textrm{d}{\bar r}_2&=&\sum_{j,j'}\sum_{n,n'}\sum_{i,i'}c_{j,n} c_{j',n'} d_{j,n,i}d_{j',n',i'} \nonumber \\ \\
		&&\times\textrm{I}({\bar r}_1,{\bar R}_j,{\bar R}_{j'},\alpha_{j,n,i},\alpha_{j',n',i'}) \nonumber,
	\end{eqnarray}
where the function $\textrm{I}({\bar r}_1,{\bar R}_j,{\bar R}_{j'},\alpha,\beta)$ is given by
	\begin{eqnarray}\label{eq:App8}
		 \textrm{I}({\bar r}_1,{\bar R}_j,{\bar R}_{j'},\alpha,\beta)&=&\frac{(4\alpha\beta)^{3/4}}{(\alpha+\beta)^{3/2}}\frac{\textrm{erf}\big(\sqrt{\alpha+\beta}|{\bar r}_1-\overline{{\bar R}}|\big)}{|{\bar r}_1-\overline{{\bar R}}|}  	
		 \nonumber \\ \\
		&&\times\exp\Bigg[-\frac{\alpha\beta\big({\bar R}_j-{\bar R}_{j'}\big)^2}{\alpha+\beta}\Bigg], \nonumber
	\end{eqnarray}
with $\overline{{\bar R}}=(\alpha{\bar R}_j+\beta{\bar R}_{j'})/(\alpha+\beta)$ and $\textrm{erf}(x)$ the error function \cite{Abramowitz}.
For the current calculation the coefficients $c_{j,n}$, $d_{j,n,i}$ and $\alpha_{j,n,i}$  are  obtained from a Hartree-Fock calculation with MOLPRO using the aug-cc-pV5Z basis set. The calculated Hartree-Fock energy for H$_{2}^{+}$ at the equilibrium distance of H$_2$, is -0.57 a.u. which is in full agreement with the exact value derived in {\cite{Wild}.

At this point a brief discussion regarding the exit point is in place.  For some simple strongly-driven atoms  the exact exit point can be computed using parabolic coordinates \cite{Parabolic}. Such computations in terms of parabolic coordinates  have been employed in a series of papers, such as \cite{Exitpoint1,Exitpoint2,Exitpoint3} to mention just a few. Recently, it was demonstrated that using the exact exit point is necessary to accurately account for the observable momentum of the tunneling electron  in attoclock experiments involving single ionization in atoms  \cite{Madsenattoclock}.  For most atoms and molecules, however, it is not possible to compute the exact exit point using parabolic coordinates.
 Thus, approximations are employed such as the one in the current work. Namely, \eq{eq:App3} is effectively a 1-d equation where only  the potential along the direction of the field is accounted for. That is,  we assume that electron 1 tunnels along the direction of the field and, using \eq{eq:App3}, we compute approximately the exit point. For the strong-field phenomena under consideration in our studies, which are double ionization and ``frustrated" double ionization,   this approximation has proven to be a very good one; our results on ``frustrated" double ionization in linearly polarized laser fields \cite{Emmanouilidou2012} are in very good agreement with experimental ones \cite{Eichmann1}.

 \subsubsection{Exit point of tunneling electron for the over-the-barrier intensity regime}\label{exitpointOVB}
If the instantaneous field strength at the initial phase $\phi_0$ is larger than the threshold  intensity for over-the-barrier ionization, then we assume that electron 1  exits in a direction opposite to the field
  at a distance $r_{max}$ \cite{Grafe}; $r_{max}$ is the coordinate  along the laser field  direction where the field-lowered Coulomb potential $V(r_{1,\parallel},t_{0})$ is maximum.
In addition, we set the magnitude of the momentum of electron 1, $\bar{p}_1$,  equal to
	\begin{equation}\label{eq:App10}
		\vert\bar{p}_1\vert=\sqrt{2(\epsilon_1-V(r_{max},t_{0}))}=\sqrt{-2(I_{p1}+V(r_{max},t_{0}))}.
	\end{equation}
The direction of $\bar{p}_1$ is uniformly distributed in space with   the only restriction being that $\bar{p}_1\cdot\bar{E}(t_0)\le0$.

\subsection{Initial phase space distribution of the nuclei}

We take the initial vibrational state of the nuclei to be the ground state of the Morse potential
 \begin{equation}
 V_M(R)=D(1-e^{-\beta(R-R_0)})^2,\label{eq:morse}
 \end{equation}
 with $R$ the internuclear distance, $D = 0.174$ a.u., $\beta = 1.029$ a.u., and $R_0$ = 1.4 a.u. (equilibrium distance of H$_2$). The relative momentum of the nuclei satisfies:
  \begin{equation}
 \frac{p_{rel}^2}{2\mu} +V_M(R)= E_0\label{eq:prel},
 \end{equation}
 where $E_0 \approx 0.01$ a.u is the vibrational ground state  and
\begin{equation}
\mu=\frac{m_{n_1}m_{n_2}}{m_{n_1}+m_{n_2}},
\end{equation}
where $m_{n_1}$ and $m_{n_2}$ are the masses of the nuclei. We choose the Wigner distribution of the ground state of the Morse potential   \cite{WignerMorse} to  describe the initial phase space distribution of the nuclei. The intensity we consider is high enough to justify restricting the initial distance of the nuclei to $R_0$ \cite{Saenz}. Concerning the relative momentum of the nuclei, $p_{rel}$, we assign to it a random number uniformly distributed in the interval [0, 10]; for values greater than 10 the Wigner distribution of the state under  consideration is essentially zero. After determining the relative momentum, we determine the momenta of the two nuclei \cite{Goldstein1980}.

Instead of the Wigner distribution we can use the classical value of the relative momentum, which we find to be equal to 4.3 a.u. from   \eq{eq:prel}. In addition, we also consider a phase space distribution with the nuclei initially at rest. We find that the Wigner and the two classical distributions yield the same results for the processes under consideration in this work, see \fig{fig:nuclei}.

 \begin{figure}[h]
  \begin{center}
    \includegraphics[width=0.85\columnwidth]{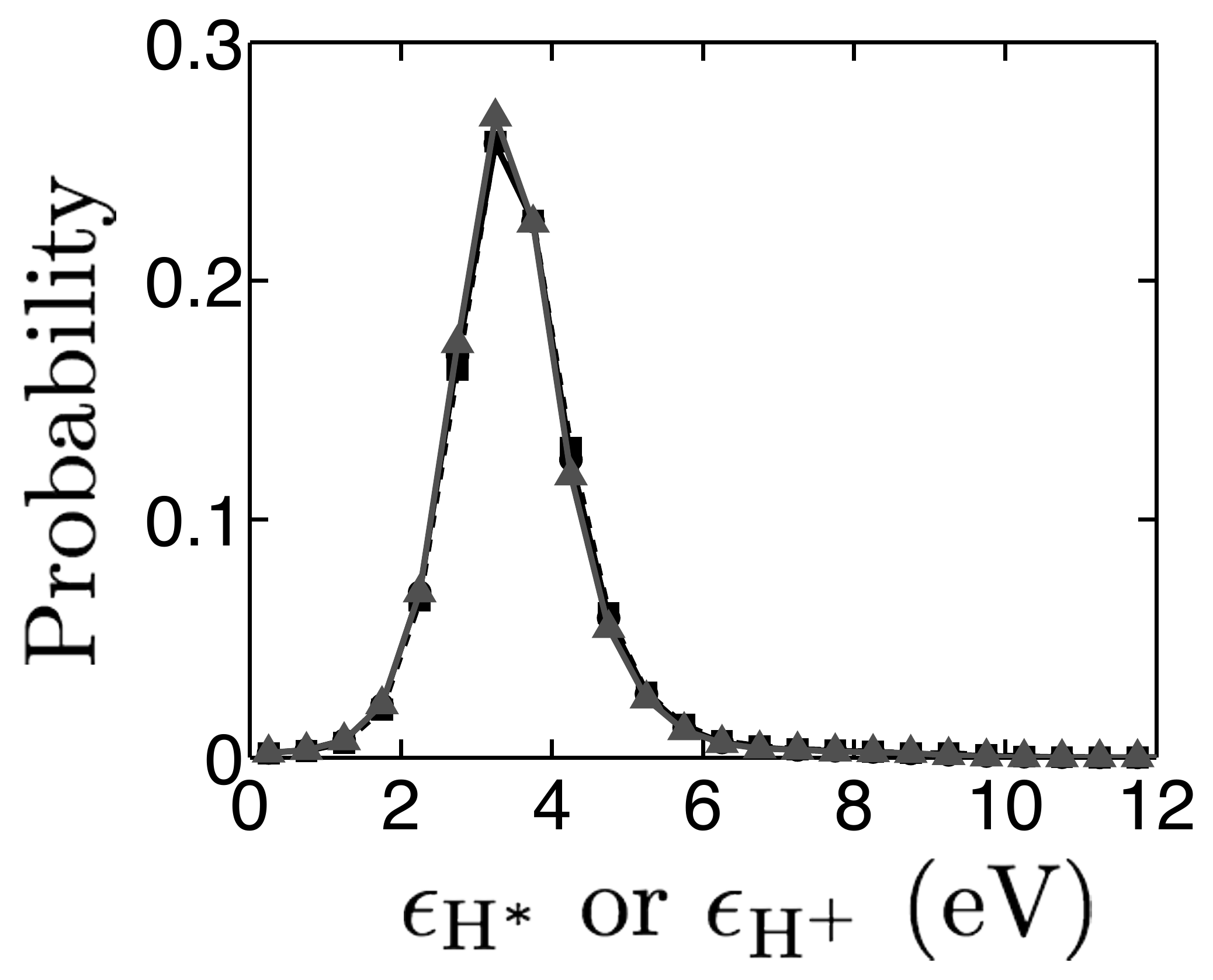}
    \caption{(Color online) The final energy distribution of the H$^+$ or H$^*$ fragments for a laser field intensity of $1.5\times10^{14}$ W$/$cm$^2$ for different initial momentum distributions of the nuclei: Wigner distribution (black solid line with full circles), 4.3 a.u. relative momentum of the nuclei (black dashed line with full squares) and nuclei initially at rest (grey solid line with full triangles).}
    \label{fig:nuclei}
  \end{center}
\end{figure}

\subsection{Propagation Technique}\label{propgationlf}
Next, we describe the technique we follow to propagate the full four-body Hamiltonian in time, i.e. including both electronic and nuclear motion. We present the technique in the context of $N$ Coulomb interacting  particles that are driven by a laser field. Previously, in \cite{Emmanouilidou2009,Emmanouilidou2012}, we formulated the equations of motion using the global regularization scheme described in \cite{Heggie1974}. In this latter work, the resulting equations of motion were propagated using the 5th order Runge-Kutta method \cite{Press2007}. In the current work, we use a time-transformed leapfrog propagation technique \cite{Mikkola2002} in conjunction with the
Bulirsch-Stoer method \cite{Stoer,Press2007}. Combining these two techniques has been used successfully to describe gravitational few-body systems  \cite{Mikkola1999, Mikkola2002, Mikkola2013}.
The advantage of the current propagation technique over the one we previously used  in \cite{Emmanouilidou2009,Emmanouilidou2012} is that it is numerically more robust with a smaller propagation error. One reason is that, unlike the technique we previously used, in the current technique the masses do not enter in the time-transformation resulting in a more accurate   treatment of many-body systems with large mass ratios \cite{Mikkola2002}.
Note that the current technique as well as the technique we previously used in \cite{Emmanouilidou2009,Emmanouilidou2012} explicitly
account for the accurate treatment of the Coulomb singularity during time propagation. This is an essential ingredient of an accurate classical treatment, since classically an electron is allowed to come infinitely close to a nucleus.

\subsubsection{Transforming to a new coordinate system}
 The Hamiltonian for $N$ Coulomb interacting particles in the presence of a laser field is given by

\begin{equation}
H=  \sum^N_{i=1}\frac{p_i^2}{2m_i}  +   \sum^{N-1}_{i=1} \sum^{N}_{j=i+1} \frac{Q_iQ_j}{|\bar{r}_i-\bar{r}_j|}   -  \sum^N_{i=1} Q_i \bar{r}_i\cdot\bar{E}(t)   ,
\label{eq:Hamiltonianleapfrog}
\end{equation}
where $Q_i$ is the charge, $m_i$ is the mass, $\bar{p}_i$ is the momentum vector and $\bar{r}_i$ is the position vector of particle $i$ and  $\bar{E}(t)= \left( E_1(t),E_2(t),E_3(t)  \right)$ is the laser-field vector. Next, we transform to a new coordinate system that involves the relative coordinate vectors $\bar{q}$ and the corresponding conjugate momenta $\bar{\rho}$, which are given by \cite{Heggie1974}

\begin{equation}
\bar{q}_{ij}=\bar{r}_i-\bar{r}_j ,
\label{eq:relativecoord}
\end{equation}

\begin{equation}
\bar{\rho}_{ij}=\frac{1}{N}\left( \bar{p}_i-\bar{p}_j-\frac{m_i-m_j}{M}\left<\bar{\rho}\right>  \right),
\end{equation}
where  $\left<\bar{\rho}\right>=  \sum^N_{i=1}\bar{p}_i$ and $M=  \sum^N_{i=1}m_i$. Expressing $\bar{r}$ and $\bar{p}$  in terms of $\bar{q}$ and $\bar{\rho}$ we obtain

\begin{equation}
\bar{r}_i=\frac{1}{M}  \sum^N_{j=i+1}m_j\bar{q}_{ij}  -\frac{1}{M}  \sum^{i-1}_{j=1}m_j\bar{q}_{ji}  +\left<\bar{q}\right> ,
\label{eq:rpostion}
\end{equation}
and

\begin{equation}
\bar{p}_i= \sum^N_{j=i+1}\bar{\rho}_{ij}  -  \sum^{i-1}_{j=1}\bar{\rho}_{ji}  +\frac{m_i}{M}\left <\bar{\rho}\right> \label{relmom}
\end{equation}
where $\left<\bar{q}\right>=\frac{1}{M}  \sum^N_{i=1}m_i\bar{r}_i$.
Next, we define a fictitious particle for each  $ij$ pair replacing the $ij$ with the $k$ index as follows

\begin{equation}
k(i,j)= (i-1)N-i(i+1)/2+j,
\end{equation}
for $i<j$ with a total of   $K=\frac{N( N-1)}{2}$ fictitious particles.  Using this notation \eq{relmom} takes the form

\begin{equation}
\bar{p}_i= \left[ \sum^K_{k=1}a_{ik}\bar{\rho}_{k} \right] +\frac{m_i}{M}\left<\bar{\rho}\right>,
\label{eq:momentumtransform}
\end{equation}
with $a_{ik}=1$ and $a_{jk}=-1$ when $k=k(i,j)$, otherwise $a_{ij}=0$.
Expressing the Hamiltonian in \eq{eq:Hamiltonianleapfrog} in terms of the relative coordinates and conjugate momenta we obtain

\begin{eqnarray}
 H &=& \sum^K_{k,k'=1} T_{kk'}\bar{\rho}_{k}\bar{\rho}_{k'} +  \frac{1}{2M}\left<\bar{\rho}\right>^2 +  \sum^{K}_{k=1} \frac{U_{k}}{q_{k}} \nonumber\\
 &-&  \left(  \sum^K_{k=1} L_{k} \bar{q}_{k}  +  \sum^N_{i=1}Q_i\left<\bar{q}\right> \right)\cdot\bar{E}(t)
\label{eq:ham1}
\end{eqnarray}
with

\begin{eqnarray}
 &&T_{kk'}=\sum^N_{i=1}  \frac{a_{ik}a_{ik'}}{2m_i},  \\
 \nonumber \\
  &&U_{k}=Q_iQ_j, \\
  \nonumber \\
  &&L_{k}  = \frac{Q_im_j - Q_jm_i}{M}
\end{eqnarray}

The equations of motion are, then, given by

\begin{eqnarray}
 \frac{d \bar{q}_{k}}{dt}= 2\sum^K_{k'=1} T_{kk'}\bar{\rho}_{k'}   && \frac{d \left<\bar{q} \right>}{dt}= \frac{1}{M}\left<\bar{\rho} \right> \\
\quad&\quad&\quad \nonumber\\
\frac{d \bar{\rho}_{k}}{dt}=   \frac{U_{k} \bar{q}_{k}}{q_{k}^3} +  L_{k} \bar{E}(t)  && \frac{d \left< \bar{\rho} \right>}{dt}=    \sum^N_{i=1}Q_i \bar{E}(t)
\label{eq:eqmotion}
\end{eqnarray}

\subsubsection{Time-transformed leapfrog}
For close encounters between two particles the Hamiltonian in \eq{eq:ham1} is singular. Previously, in \cite{Emmanouilidou2009,Emmanouilidou2012}, this issue was addressed by transforming to regularized coordinates \cite{Heggie1974}. In the current work, to address the singularity, we use the time-transformed leapfrog method that is described in \cite{Mikkola2002}; we can do so, since in \eq{eq:eqmotion} the derivative expressions are independent of the quantities themselves. In the leapfrog method two sets of first order differential equations are identified. In our case, these two sets correspond to the relative coordinates $\bar{q}$ and the corresponding conjugate momenta  $\bar{\rho}$.  In addition, we consider the time transform $ds = \Omega(\bar{q}) dt$  \cite{Mikkola2002}; $\Omega(\bar{q})$  is an arbitrary positive function of the relative position vectors. Introducing a new auxiliary variable $W=\Omega$ the equations of motion take the form $\bar{q}'=\dot{\bar{q}}(\bar{\rho})/W$, $t'=1/W$ and $\bar{ \rho}'=\dot{\bar{\rho}}(\bar{q})/\Omega$; prime denotes  the derivate with respect to the new time variable $s$.
Instead of using the relation $W=\Omega$ directly, we obtain the value of $W$ from the differential equation:

\begin{equation}
\frac{d W}{dt} =  \dot{\bar{q}}(\bar{\rho}) \cdot \frac{\partial\Omega(\bar{q})}{\partial \bar{q}}.
\end{equation}
Applying the leapfrog method we now propagate $\bar{q}$, $t$, $\bar{\rho}$ and $W$ over a time-step $h$ as follows: i) we propagate $ \bar{q}$ and  $t$ over half a time-step, $h/2$; ii)  we propagate   ${\bar{\rho}}$ and $W$ over a time-step $h$ using the values of $ \bar{q}$ and $t$ at  half the time step $h/2$. For each pair of a relative coordinate $\bar{q}$ and the corresponding conjugate momentum  $\bar{\rho}$ the time-transformed leapfrog set of equations take the form:

\begin{equation}
 \begin{array}{l}
\bar{q}_{1/2}= \bar{q}_{0}+\frac{h}{2}\frac{\dot{\bar{q}}( \bar{\rho}_{0})}{W_0} \\
 t_{1/2}= t_{0}+\frac{h}{2}\frac{1}{W_0} \\
\quad \\
\bar{\rho}_{1}= \bar{\rho}_{0}+h \frac{\dot{\bar{\rho}}(\bar{q}_{1/2})}{\Omega(\bar{q}_{1/2})} \\
 W_{1}=  W_{0}+h \frac{   \dot{\bar{q}}(\bar{\rho}_{0}) +  \dot{\bar{q}}(\bar{\rho}_{1})  }{2\Omega(\bar{q}_{1/2})} \cdot \left. \frac{\partial\Omega(\bar{q})}{\partial \bar{q}} \right |_{\bar{q}=\bar{q}_{1/2}}  \\
\quad \\
\bar{q}_{1}= \bar{q}_{1/2}+\frac{h}{2}\frac{\dot{\bar{q}}(\bar{\rho_{1}})}{W_1} \\
 t_{1}=  t_{1/2}+\frac{h}{2}\frac{1}{W_1} \\
 \end{array}
\end{equation}
where the subscripts 0, 1/2, 1 denote the values of the variables at the initial time, at half a time-step and at the end of a time-step. Note that we have $K$ such sets of equations, as many as the number of fictitious particles. We choose   $\Omega$ so that   if any of the relative coordinates becomes small (two-body close encounter) then the  time-step reduces accordingly:

\begin{equation}
\Omega =  \sum^K_{k=1} \frac{1}{\left |\bar{q}_{k}\right |}.
\end{equation}

\subsubsection{Bulirsch-Stoer Method}

The final step in the integration of the equations of motion, involves incorporating the leapfrog method into the Bulirsch-Stoer method \cite{Press2007,Stoer}. In this latter method, the propagation over a time step $H$ takes place by splitting it into $n$ substeps of size $h=H/n$. For the propagation over each one of these substeps, we use the time-transformed leapfrog technique. The algorithm we follow  to propagate is given by  \cite{Mikkola1999,Mikkola2013}

\begin{equation}
\begin{array}{l}
\bar{q}_{1/2}= \bar{q}_{0}+\frac{h}{2}  \frac {\dot{\bar{q}}(\bar{\rho}_{0})}{W_0} \\
t_{1/2}= t_{0}+\frac{h}{2}\frac{1}{W_0}\\
\quad \\
\bar{\rho}_{1}= \bar{\rho}_{0}+ h \frac{\dot{\bar{\rho}}(\bar{q}_{1/2})}{\Omega(\bar{q}_{1/2})} \\
W_{1}=  W_{0}+h \frac{   \dot{\bar{q}}(\bar{\rho}_{0}) +  \dot{\bar{q}}(\bar{\rho}_{1})  }{2\Omega(\bar{q}_{1/2})} \cdot \left. \frac{\partial\Omega(\bar{q})}{\partial \bar{q}} \right |_{\bar{q}=\bar{q}_{1/2}}  \\
\quad \\
\bar{q}_{m-1/2}= \bar{q}_{m-3/2}+h \frac{  \dot{\bar{q}}(\bar{\rho}_{m-1})}{W_{m-1}} \\
t_{m-1/2}= t_{m-3/2}+h\frac{1}{W_{m-1}} \\
\vdots  \\
\bar{\rho}_{m}= \bar{\rho}_{m-1}+h  \frac{   \dot{\bar{\rho}}(\bar{q}_{m-1/2}) }{\Omega(\bar{q}_{m-1/2})}  \\
 W_{m}=  W_{m-1}+h \frac{   \dot{\bar{q}}(\bar{\rho}_{m-1}) +  \dot{\bar{q}}(\bar{\rho}_{m})  }{2\Omega(\bar{q}_{m-1/2})} \cdot \left. \frac{\partial\Omega(\bar{q})}{\partial \bar{q}} \right |_{\bar{q}=\bar{q}_{m-1/2}}\\
\vdots \\
\bar{q}_{n}= \bar{q}_{n-1/2}+\frac{h}{2} \frac{ \dot{\bar{q}}(\bar{\rho}_{n}) }{W_n}  \\
t_{n}= t_{n-1/2}+\frac{h}{2}\frac{1}{W_n} \\
\label{eq:leapfrogexample}
 \end{array}
\end{equation}
where $m=2,...,n$. This process of integrating from $\bar{q}_{0}$, $\bar{\rho}_{0}$ to $\bar{q}_{n}$, $\bar{\rho}_{n}$  is repeated with  increasing values of $n$ until  extrapolation to zero time-step, i.e.  $\bar{q}_{n}$ and $\bar{\rho}_{n}$ for $n\rightarrow \infty$, is achieved with satisfactory error.
Using the techniques described above we obtain results similar to those in \cite{Emmanouilidou2012} for H$_2$ when driven by a linearly polarized laser field. The current technique is numerically more robust than the one used in \cite{Emmanouilidou2012} and we, thus, adopt it in what follows.

\subsection{Tunneling during propagation}

  During time propagation, we allow each electron  to tunnel
at the classical turning points along the field axis using the Wentzel-Kramers-Brillouin (WKB) approximation, for details see \cite{Cohen}.
 For the transmission probability  we use the WKB formula for transmission through a potential barrier \cite{Merzbacher}
	\begin{equation}\label{eq:App20}
		T\approx\exp\left(-2\int_{r_{a}}^{r_{b}}(2(V_{tun}(r,t_{tun})-\epsilon_{n}))^{1/2}\textrm{d}r\right),
	\end{equation}
with $\mathrm{V_{tun}(r, t_{tun})}$ the potential along the field direction of each electron in the presence of the nuclei and the laser field, which is of the same form as the potential in \eq{eq:App3} except for the integral term; $\mathrm{\epsilon_{n}}$ is the energy of an electron at the time of tunneling, $t_{tun}$, and $r_{a}$ and $r_{b}$ are the classical turning points. We find that accounting for tunneling during time-propagation is necessary in order
to accurately describe phenomena related to enhanced ionization during  the fragmentation of strongly-driven molecules.

\subsection{Identifying Rydberg states in neutral atoms}
In what follows, we adopt a Classical Trajectory Monte Carlo (CTMC) method that involves all the techniques discussed in the previous sub-sections. We use this CTMC method to describe the formation of highly excited neutral atoms, through Coulomb explosion, in strongly-driven H$_2$. After propagating the trajectories to the asymptotic limit we select trajectories that produce, H$^{+}$, a free electron and H$^{*}$ (where $*$ denotes that the electron is in a $n>1$ quantum state). To identify the trajectories when the electron is captured in an excited state, we first find the classical principal number $n_{c}=1/\sqrt{2| \epsilon_{n} |}$, where $\epsilon_{n}$ is the total energy of the trapped electron.  We, next, assign a quantum number so that the following criterion,  which is derived in \cite{Mackellar}, is satisfied:

\begin{equation}
[(n-1)(n-1/2)n]^{1/3} \leq n_c \leq [n(n+1/2)(n+1)]^{1/3}.
\end{equation}

\section{Results}
In what follows we consider two laser field intensities, specifically,  $1.5\times10^{14}$ W$/$cm$^2$ in the tunneling regime and $2.5\times10^{14}$ W$/$cm$^2$ in the over-the-barrier regime.
In \fig{fig:3} we compute   the distribution of the  quantum number $n$ for  $\epsilon=0$ and  $\epsilon=0.45$ for the two field intensities. We find that the $n$ quantum number peaks around 8 in all cases considered. In \fig{fig:2} we show the energy distribution of the H$^+$ and H$^*$ fragments for the same two intensities and ellipticities of the laser field.  The energy distribution of the H$^+$ and H$^*$ fragments remains roughly the same as a function of ellipticity while it peaks at a slightly higher value for $2.5\times10^{14}$ W$/$cm$^2$ compared to $1.5\times10^{14}$ W$/$cm$^2$.
\begin{figure}[h]
  \begin{center}
    \includegraphics[width=0.85\columnwidth]{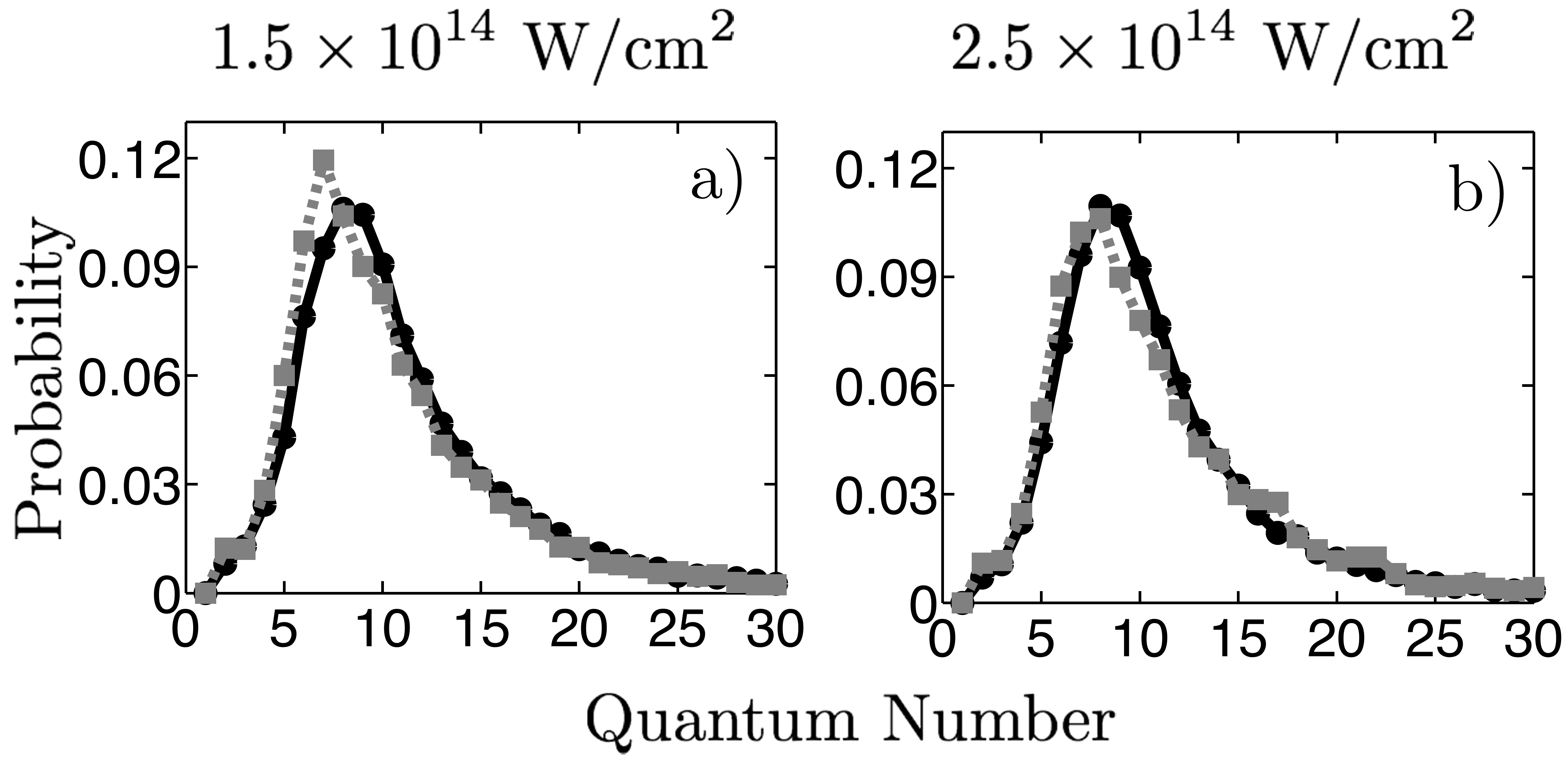}
    \caption{(Color online) The distribution of the quantum number $n$ for a field intensity $1.5\times10^{14}$ W$/$cm$^2$ a), and $2.5\times10^{14}$ W$/$cm$^2$ b). The black solid line with full circles corresponds to
    $\epsilon=0$, and the grey dashed line with full squares  corresponds to $\epsilon=0.45$. }
    \label{fig:3}
  \end{center}
\end{figure}

\begin{figure}[h]
  \begin{center}
    \includegraphics[width=0.85\columnwidth]{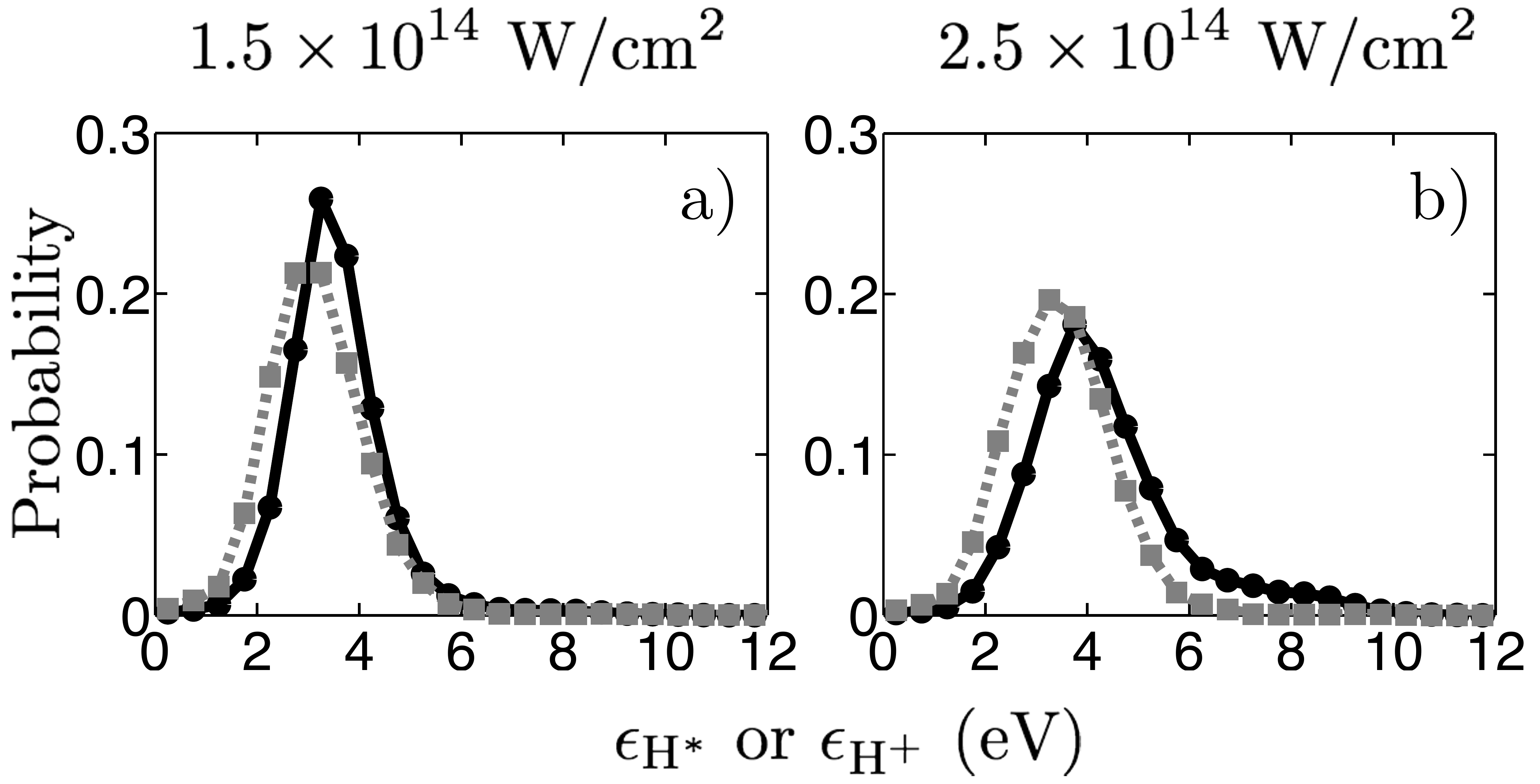}
    \caption{(Color online) The final energy distribution of the H$^+$ or H$^*$ fragments for a field intensity $1.5\times10^{14}$ W$/$cm$^2$ a), and $2.5\times10^{14}$ W$/$cm$^2$ b). The black solid line with full circles corresponds to    $\epsilon=0$, and the grey dashed line with full squares corresponds to $\epsilon=0.45$. }
    \label{fig:2}
  \end{center}
\end{figure}

Next, we  investigate the dependence of the two pathways of  $\mathrm{H^{*}}$ formation on the degree of ellipticity of the laser field.
 These pathways  can be separated as to which one of the two ionization steps, i.e., the earlier tunnel ionization of electron 1  or the later tunnel ionization of electron 2  is ``frustrated"  \cite{Emmanouilidou2012}.   In \fig{fig:cartoon} a) we show pathway A where electron 1 tunnel ionizes, subsequently escaping very quickly. Electron 2, later, tunnel ionizes and quivers in the laser field; however, when the field is turned off, electron 2 does not have enough drift energy to escape and occupies a Rydberg state of the H-atom instead. Hence, in Pathway A the later ionization step is ``frustrated".
In \fig{fig:cartoon} b) we show pathway B where electron 1 tunnel ionizes very quickly, quivering in the field, while electron 2 tunnel ionizes and escapes after a few periods of the laser field. When the laser field is turned off, electron 1 does not have enough energy to escape and remains in a Rydberg state of the H-atom instead, i.e., the earlier ionization step is ``frustrated".

 \begin{figure}[h]
\centerline{\includegraphics[scale=0.165,clip=true]{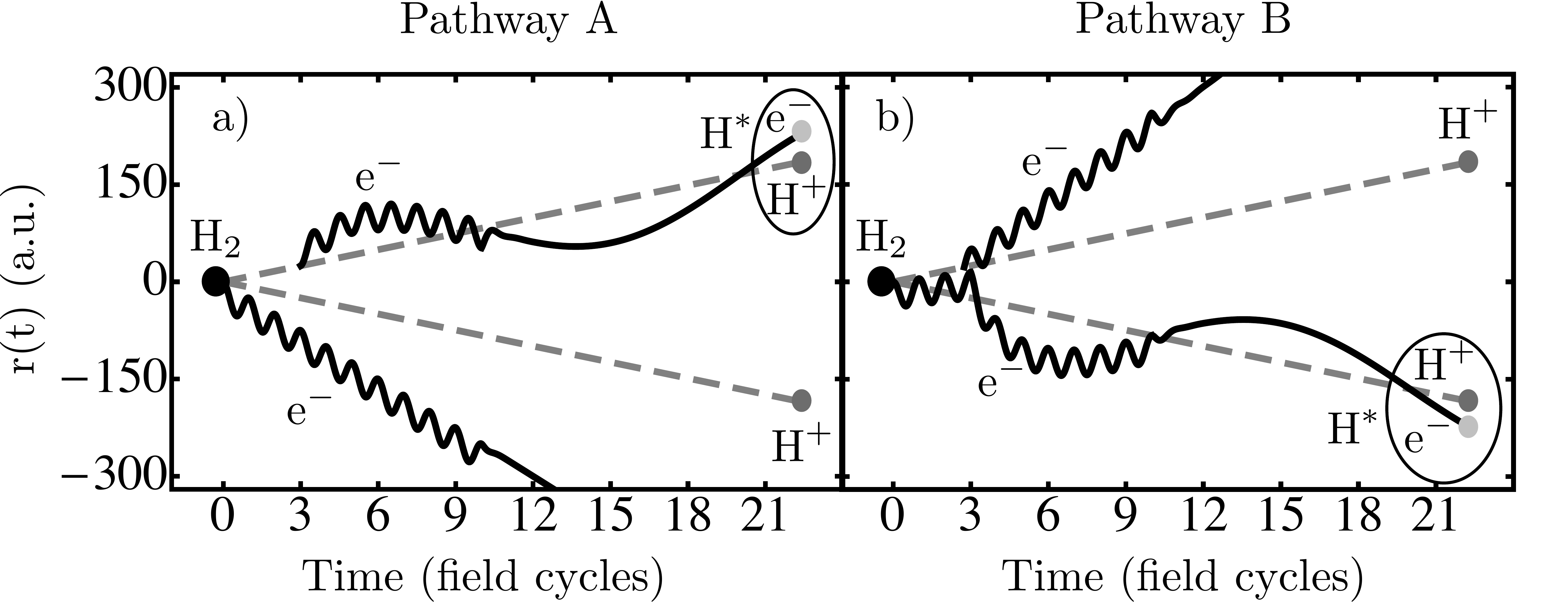}}
\caption{\label{fig:cartoon}
(Color online) Schematic illustration of the two routes leading to formation of $\mathrm{H^{*}}$ for $\epsilon=0$: a) Pathway A, b) Pathway B. Shown is the time-dependent position along the laser field for electrons (black lines) and ions (gray broken  lines). This figure appears in \cite{Emmanouilidou2012}; we also include it here  for completeness.}
\end{figure}

In \fig{fig:1} we show how the probability of pathway A and B (out of all trajectories) changes with the degree of ellipticity of the laser field. For the smaller intensity of $1.5\times10^{14}$ W$/$cm$^2$, we find that as $\mathrm{\epsilon}$ increases the probability of pathway B drops more sharply than that of A.
For instance, for  $\epsilon=0$ pathway B is 1.6 times more probable than pathway A, while for $\epsilon=0.45$ pathway B is roughly 6 times less probable than  A. Thus, for  the smaller intensity in the tunneling regime, for $\epsilon>0.4$ pathway B is practically ``switched-off" with pathway A prevailing. For the higher  intensity of $2.5\times10^{14}$ W$/$cm$^2$, we find that as $\epsilon$ increases the probability of pathway B drops even more sharply compared to the smaller intensity. For instance,   for  $\epsilon=0$ pathway B is roughly as probable  as pathway A, while for $\epsilon=0.45$ pathway B is roughly 25 times less probable than  A. Thus, for the higher intensity in the over-the-barrier regime, for  $\epsilon>0.3$ pathway B is practically ``switched-off".

\begin{figure}
  \begin{center}
    \includegraphics[width=0.85\columnwidth]{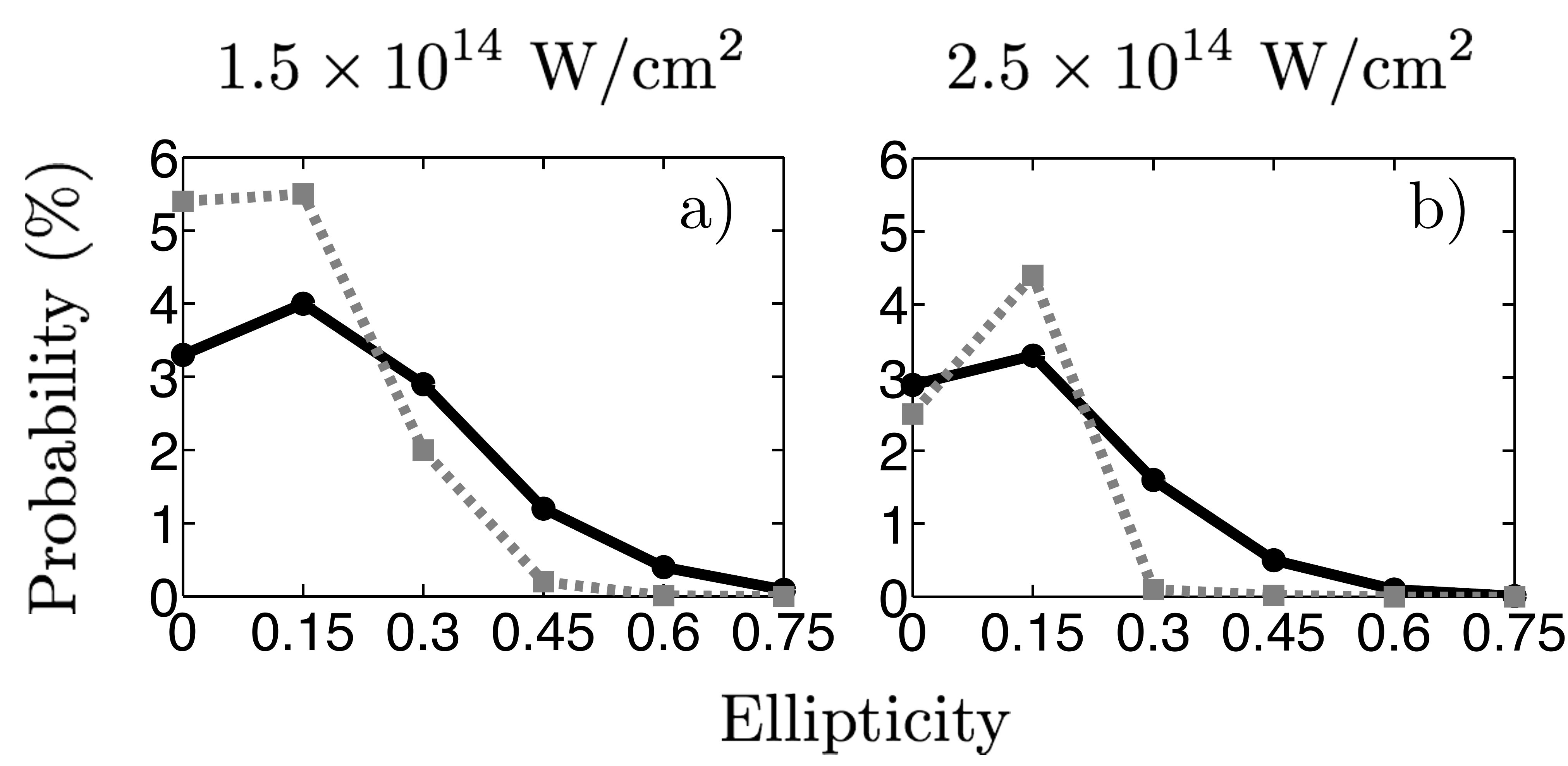}
    \caption{(Color online) Probabilities for the two pathways for an intensity $1.5\times10^{14}$ W$/$cm$^2$ a), and $2.5\times10^{14}$ W$/$cm$^2$ b). The black solid line with full circles is for
    pathway A, and the grey dashed line with full squares is for pathway B. }
    \label{fig:1}
  \end{center}
\end{figure}

The question that naturally arises is why pathway B is more sensitive to the ellipticity of the laser field. Double ionization  events where re-collisions prevail are very sensitive to $\epsilon$. The reason is that
 a slight ellipticity of the laser field offsets the electron from the ion roughly by
$5\epsilon E_{0}/\omega^2$ making a re-collision less probable \cite{Dietrich1996}. The sensitivity to ellipticity of our  ``frustrated" double ionization events for pathway B
strongly suggests that two-electron effects in the form of re-collisions underlie pathway B and not  pathway A. This explanation is also consistent with pathway B being ``switched-off"
faster for  $2.5\times10^{14}$ W$/$cm$^2$ than for $1.5\times10^{14}$ W$/$cm$^2$: the offset of the re-colliding electron  from the ion core increases with increasing intensity of the laser field.

Indeed, in \cite{Emmanouilidou2012} we have provided evidence that one-electron effects prevail in pathway A, while two-electron effects prevail in pathway B. That is, we have shown that in pathway A electron 2 transitions from the ground state of the $\mathrm{H_{2}}$ molecule to a high Rydberg state of the H-atom by gaining energy through a strong interaction with the laser field. This gain of energy resembles enhanced ionization in $\mathrm{H_{2}^{+}}$  \cite{bandrauk1996}. We have also provided evidence that in pathway B electron 2 gains energy  to ionize mainly through
two-electron effects resembling Delayed NSDI (non-sequential double ionization)
which is a major pathway of double electron escape (also referred to as re-collision-induced excitation with subsequent field ionization, RESI \cite{Delayed}). In Delayed NSDI (weak re-collision) the re-colliding electron returns to the core close to a zero of the field, transfers energy to the second electron and one electron escapes with a delay after re-collision.  For pathway B the electron-electron correlation  is in the form of  ``frustrated" delayed NSDI since one electron eventually does not escape. From the above,  it follows that the dependence  of the probability of pathways A and B  on $\epsilon$ (\fig{fig:1}) provides strong support that re-collisions underlie pathway B.

Pathway A also decreases with ellipticity, even though pathway A is less  sensitive to ellipticity compared to pathway B. To understand this decrease we consider the change in momentum, due to the laser field, of the electron that tunnel ionizes in pathway A, i.e. of electron 2. This change   is roughly $2\sqrt{U_p}(\sin(\omega t_{tun})\hat{z}-\epsilon \cos(\omega t_{tun})\hat{x})$, where $t_{tun}$ is the time of  tunnel ionization and $U_{p}=E_0^2/4\omega^2$. Moreover, since tunnel ionization takes place mostly around a maximum of the laser field,
the change in momentum  of electron 2 reduces to $2\epsilon\sqrt{U_p}\hat{x}$. Thus, with increasing ellipticity the  momentum of electron 2 increases. As a  result ``frustrated" double ionization
events are converted  to double ionization events, accounting for the decrease with ellipticity of the probability of pathway A.

\fig{fig:1} shows that  two-electron effects are essentially ``switched-off'' in $\mathrm{H^{*}}$ formation for $\epsilon>0.4$ for an intensity $1.5\times10^{14}$ W$/$cm$^2$ and for $\epsilon>0.3$ for an intensity $2.5\times10^{14}$ W$/$cm$^2$ with pathway A prevailing. This prevalence of one-electron effects with increasing $\epsilon$ is also evident in the observable momentum space of the escaping electron.
 In \fig{fig:4} we plot the total x-z  momentum distribution  of the escaping electron  for ellipticities 0 and 0.45 for the two laser field intensities currently under consideration.  The total 2-d distributions account for both pathways and all initial tunneling directions of electron 1. For $\epsilon=0$ (\fig{fig:4} a) and c)) the traces of both pathways A and B (\fig{fig:1}) are present in the 2-d momentum distributions.   The trace of pathway B  is the large spread in momentum \cite{Emmanouilidou2012} which is mostly due to the strong interaction of electron 2 with the Coulomb potential \cite{Comtois}. However, for larger values of $\epsilon$ this large spread disappears, see \fig{fig:4} b) and d); this is a clear signature of the prevalence of pathway A. Note that
 with increasing ellipticity the highest momentum along the x-axis increases. This is expected since  the maximum  change in momentum along the x-axis, due to the laser field,  is approximately given by $2\epsilon\sqrt{U_p}$.  We note that the 2-d momentum distributions for the higher intensity reach higher values of momentum in $\sqrt{U_{p}}$ than for the lower intensity. The reason, most probably, is that while in the tunneling regime electron 1 tunnel ionizes at time zero with zero velocity along the direction of the laser field,  in the over-the-barrier regime electron 1 tunnels with non zero velocity.

 Moreover,  for larger values of $\epsilon$, see \fig{fig:4} b) and d), we obtain   an asymmetric two-lobe momentum distribution. This  asymmetry, first observed in \cite{ Coulombasymmetry1}, has sparked a lot of studies in single ionization of atoms in elliptical fields. It  has been, mainly,  attributed to the effect of the Coulomb potential  \cite{Coulombasymmetry2}. Since our 3-d semiclassical model fully accounts for the Coulomb potential the asymmetry in the momentum distribution is also evident in our results in \fig{fig:4} b) and c).
Besides the current study, studies of this asymmetry for molecular systems are few; they include a theoretical one of  strongly-driven $\mathrm{H_{2}^{+}}$ \cite{Grafe} and an experimental one on double ionization of $\mathrm{H_{2}}$ \cite{Andre}.
 Our results for $\mathrm{H^{*}}$ formation in \fig{fig:4} b) and d) show that with increasing $\epsilon$ the two-lobe structure tends to align closer to the minor axis of the field (x-axis in our case) \cite{Andre, Grafe}.

\begin{figure}
  \begin{center}
    \includegraphics[width=0.85\columnwidth]{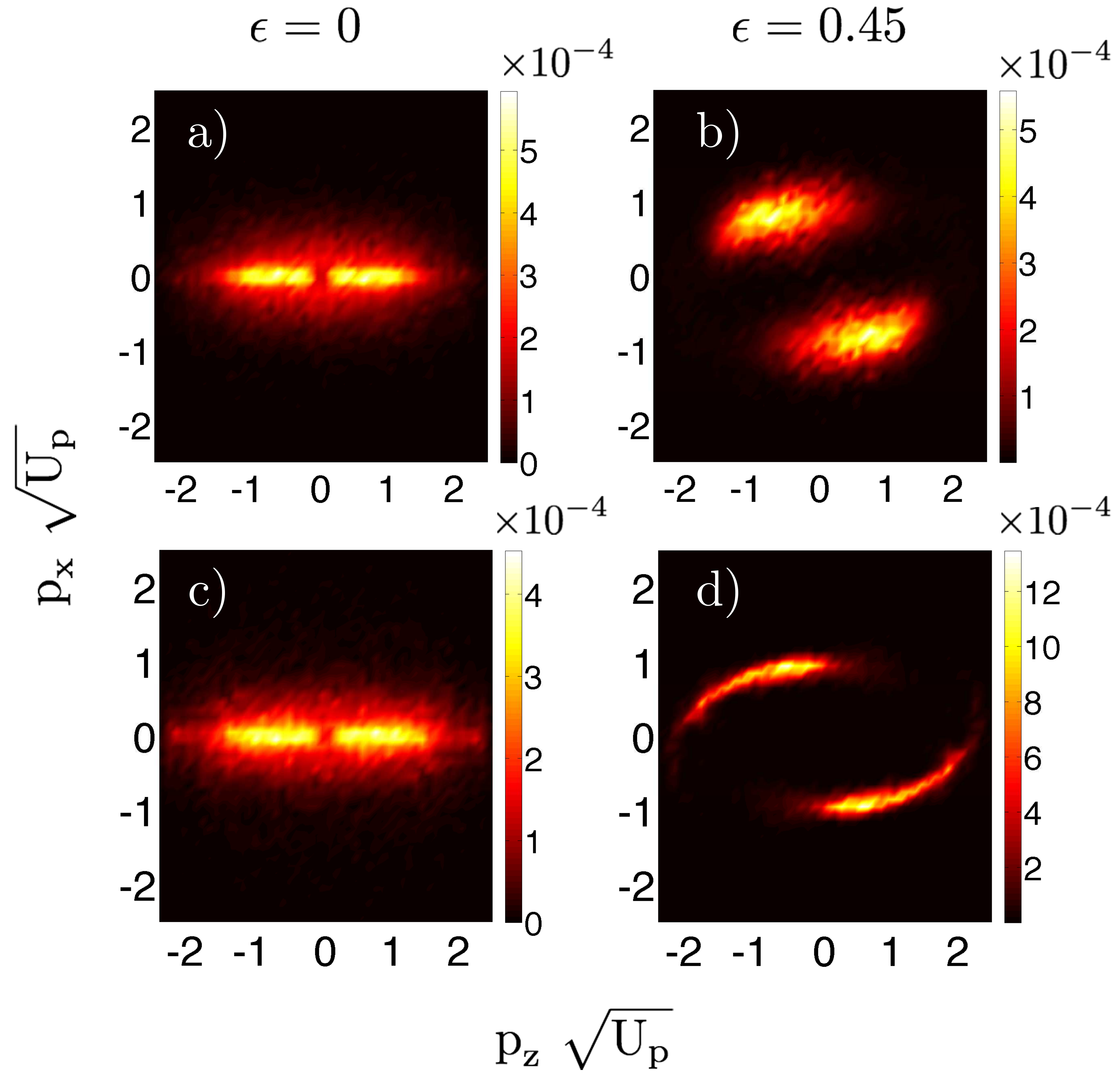}
    \caption{(Color online) The 2-d electron momentum distribution for an intensity  $1.5\times10^{14}$ W$/$cm$^2$ a) and b), and for $2.5\times10^{14}$ W$/$cm$^2$ c) and d). Figures on the left are for $\epsilon=0$, and those on the right are for $\epsilon=0.45$. The momentum is expressed in $\sqrt{\textrm{U}_\textrm{p}}$.}
    \label{fig:4}
  \end{center}
\end{figure}

Finally, we briefly discuss  the sites electron 2  tunnel ionizes from.
 Specifically,  we consider the combined potential of electron 2 in the presence of the two nuclei and the laser field along the direction of the laser field (tunneling direction).
 Our results indicate that when electron 2 tunnel ionizes the inter-nuclear distances   range from intermediate to large. For these  distances and at times close to extrema of the field we find that the  potential of electron 2 along the direction of the field has either a  double or a single-well. For the double-well an inner barrier is present such that the potential of electron 2 is higher in one well (up-field)
compared to the other well (low-field) (as is the case for enhanced ionization \cite{bandrauk1996,Niikura,Wu2012}).  The tunnel ionization sites are thus an up-field, low-field and a single-well, see \fig{fig:sites}. We consider ellipticities up to 0.45.  For $1.5\times10^{14}$ W$/$cm$^2$ we find that, out of all  H$^{*}$ formation events,  electron 2 tunnel ionizes from an up-field well in 85\% of the cases while  from a low-field or a single-well in 10\% of the cases.
For  $2.5\times10^{14}$ W$/$cm$^2$, we find that electron 2 again mostly tunnel ionizes from an up-field well, however, there is an increased probability to tunnel ionize from a low-field or a single well compared to the lower intensity. 

\begin{figure}[h]
  \begin{center}
    \includegraphics[width=0.7\columnwidth]{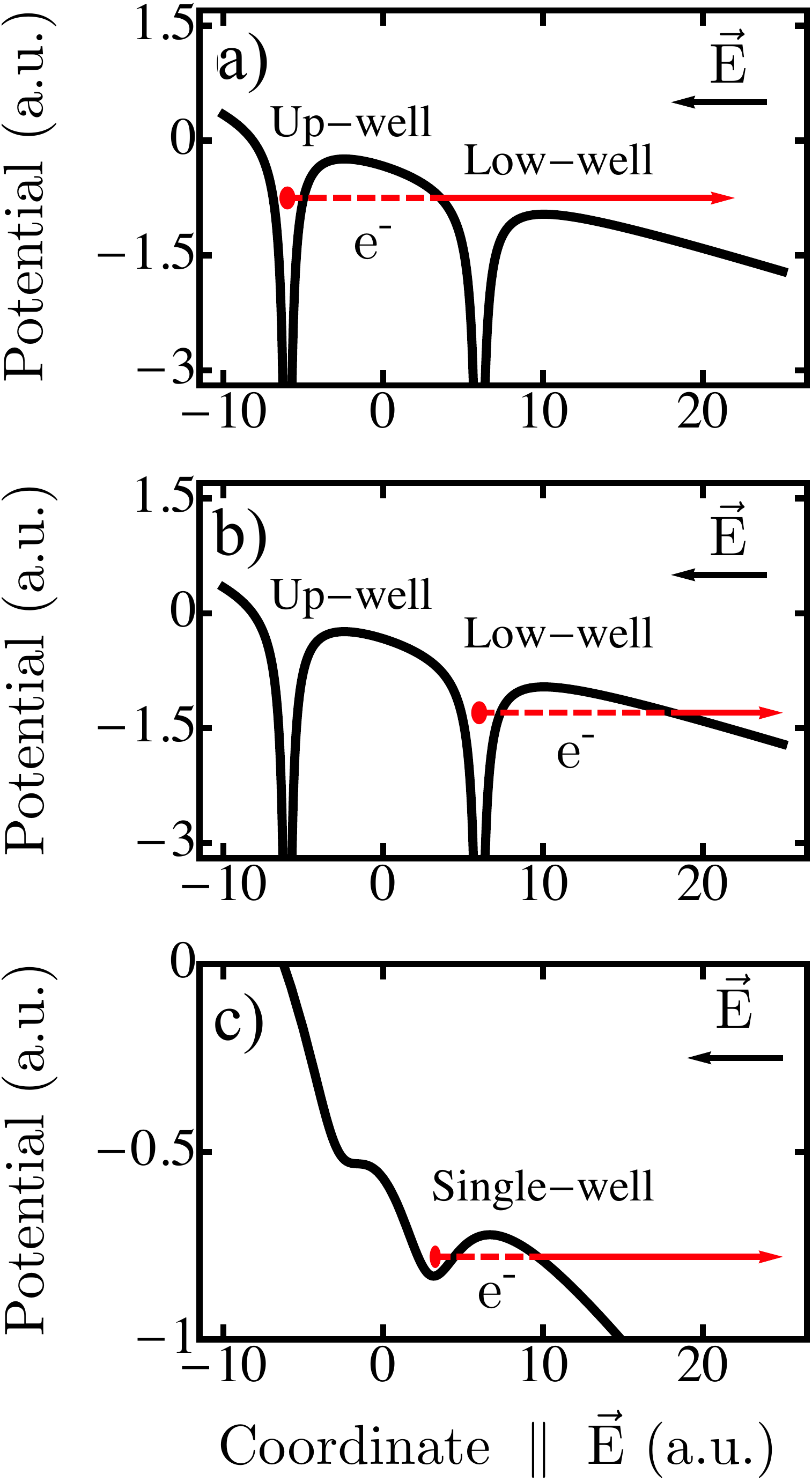}
    \caption{(Color online) Schematic illustration of the sites an electron can tunnel from: a) up-field well, b)  low-field well and c) single-well.   }
    \label{fig:sites}
  \end{center}
\end{figure}

\section{Conclusions}
We have presented a toolkit for semiclassical computations of strongly-driven molecules.
This toolkit includes the formulation  of the probabilities of strong-field phenomena  in a transparent way. This formulation allowed us to identify an electronic initial phase space distribution for the over-the-barrier intensity regime that works better than previous ones. However, more work is needed to formulate more accurate initial phase  space distributions for the electronic degrees of freedom
in the over-the-barrier regime. This toolkit also includes a 3-dimensional method for time-propagation that fully accounts for the Coulomb singularity. This 3-d method combines the time-transformed leapfrog propagation technique and the Bulirsch-Stoer method and has been previously developed in the context of celestial mechanics. In the current work, we adopt  this technique to strongly-driven systems. Another important element of this toolkit is allowing for tunneling during  propagation, that is, the time-propagation is not classical. We find that the latter is necessary in order to accurately
describe phenomena associated with enhanced ionization in the fragmentation of  strongly-driven  molecules.
In the current work, using this toolkit, we elucidated the interplay   of the electronic and nuclear dynamics in $\mathrm{H^{*}}$ formation during the break-up of strongly-driven $\mathrm{H_{2}}$ by elliptical laser fields.  We find that with increasing ellipticity  we  ``switch-off" two-electron effects. That is, we  find that pathway A, which is similar to a  ``frustrated" enhanced ionization process, prevails. Moreover, we have shown that the observable momentum space of the escaping electron clearly bears the imprints of one-electron effects with increasing ellipticity.

{\it Acknowledgments.} A.E. acknowledges helpful discussions with Dr. Andre Staudte and support from the EPSRC grants no. H0031771 and J0171831 and the use of the computational resources of Legion at UCL.

\appendix
\section{Ionization rate}\label{ionrate}

For both the below- and the over-the-barrier intensity regimes we use a semiclassical formula for the tunneling rate that was derived  in \cite{Murray2011}

\begin{equation}\label{eq:App12}
		\Gamma=2\pi \kappa^2C^2_\kappa\Bigg(\frac{2\kappa^3}{\vert\bar{E}(t_0)\vert}\Bigg)^{2Q/\kappa-1}\exp\Bigg(-\frac{2\kappa^3}{3\vert\bar{E}(t_0)\vert}\Bigg)R(\theta_L),
	\end{equation}
where $\vert\bar{E}(t_0)\vert$ is the instantaneous field strength, $\theta_L$ is the angle between the laser field and the z-axis in the molecular frame, $\kappa=\sqrt{2I_{p1}}$, and $Q$ is the asymptotic charge. For H$_2$ the asymptotic charge is equal to one.
The coefficient $C_\kappa$ is obtained by fitting the Dyson orbital to the following asymptotic form of the wave function
	\begin{equation}\label{eq:App13}
		 \Psi(\bar{r})\approx C_\kappa\kappa^{3/2}(r\kappa)^{Q/\kappa-1}e^{-\kappa r}F(\cos\theta,\sin\theta\cos\phi).
	\end{equation}
The Dyson orbital \cite{Patchkovskii2007} is the overlap integral of the two-electron wave function of the molecule with the one-electron wave function of the molecular ion; for the current work the overlap integral is that of
the ground state of $\mathrm{H_{2}}$ with the 1$\sigma_g$ state of $\mathrm{H_{2}^+}$ computed at the equilibrium distance of $\mathrm{H_{2}}$. We derive both wave functions with the Hartree-Fock method, using MOLPRO  \cite{MOLPRO_brief}.  For H$_2$ the Hartree-Fock energy obtained is -1.134 a.u., which has a 3.5\% relative difference from the experimental value  of -1.175 a.u. \cite{H2energy}.

\begin{figure}[h]
  \begin{center}
    \includegraphics[width=0.85\columnwidth]{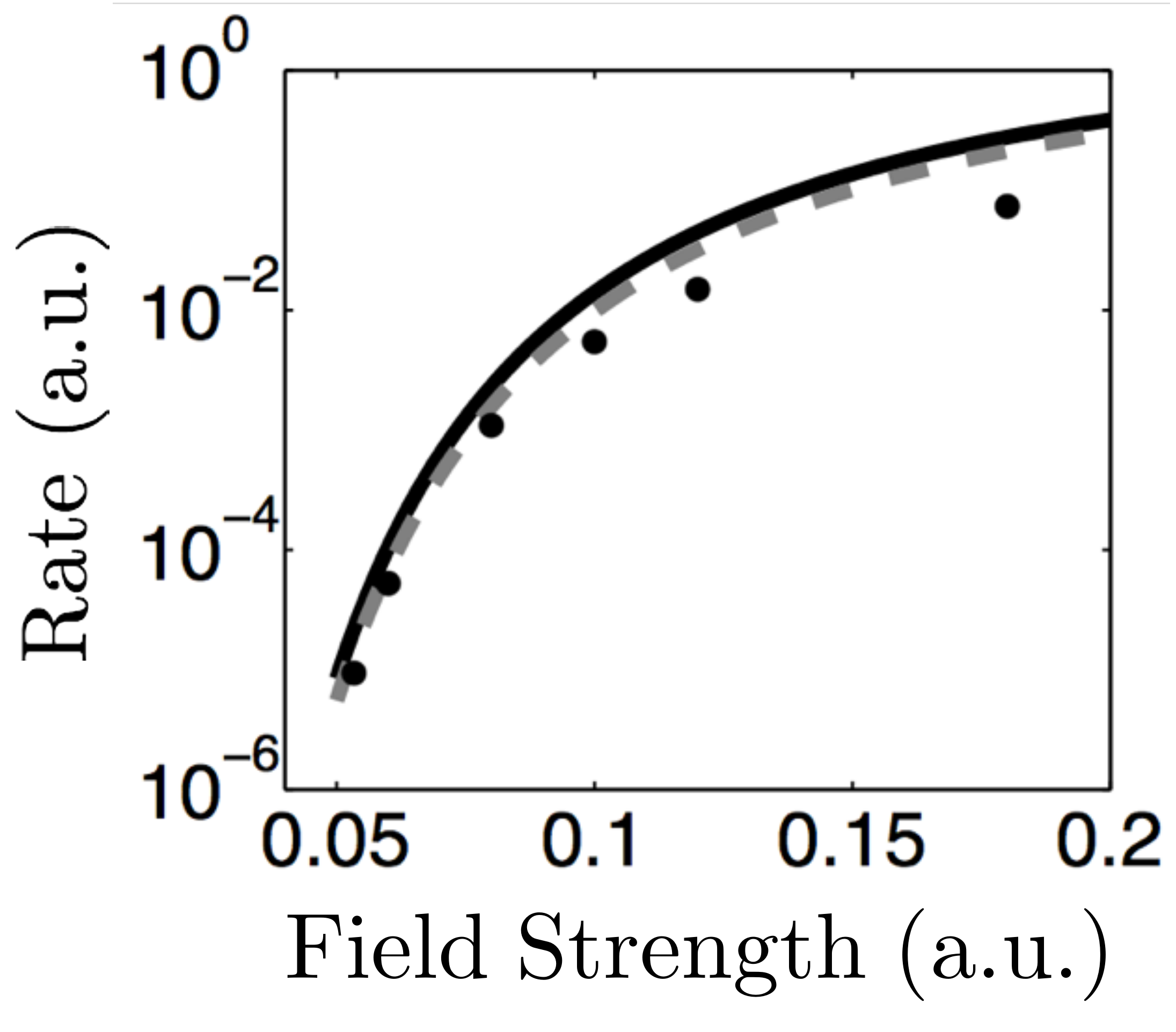}
    \caption{(Color online) a) The ionization rate of H$_{2}$ versus the field strength for a laser field parallel to the molecular axis, calculated with Eq.~(\ref{eq:App12}) (black solid), obtained from \cite{CDLin2010} (grey dashed), and obtained from \cite{Saenz2002} (full circle). }
    \label{fig:IonizationRate}
  \end{center}
\end{figure}

The function $\mathrm{F(\cos\theta,\sin\theta\cos\phi)}$ depends on the molecular orbital the electron occupies before tunneling.  For H$_2$ the electron occupies a $1\sigma_g$ orbital \cite{Levine},which we can approximately express as a LCAO of two 1s orbitals
\begin{equation}\label{eq:orbital1}
	\Phi_{1\sigma_g}(\bar{r})\propto e^{-\kappa\vert\bar{r}-\bar{R}_1\vert}+e^{-\kappa\vert\bar{r}-\bar{R}_2\vert}.
\end{equation}
Taking the asymptotic expansion for $r\gg R_0$, we derive an expression for $F(\cos\theta,\sin\theta\cos\phi)$
\begin{equation}\label{eq:orbital2}
	F(\cos\theta)=\cosh\Bigg(\frac{\kappa R_0}{2}\cos\theta\Bigg),
\end{equation}
with 	$\mathrm{R_0}$ the distance between the nuclei. An alternative expression is provided in \cite{Radzig}
	\begin{equation}\label{eq:App14}
		F(\cos\theta)=\cosh\Bigg(\frac{\kappa R_0}{2}\cos\theta\Bigg)[1+\alpha\cos^2\theta].
	\end{equation}
We find that both expressions give similar results for the tunneling rate. After fitting the Dyson orbital in the interval $3\le r\le 6$ a.u. and $0\le\theta\le\pi$ we find $C_\kappa=0.51$ and $\alpha=5.4\times10^{-3}$ for $\mathrm{H_{2}}$. The interval was chosen so that for $r>3$ a.u., the Coulomb potential corresponding to the H$_2^+$ molecular ion has effectively the form of a one-center Coulomb potential, i.e. $-Q/r$; the upper limit was chosen so that  for $r>6$ a.u. the Dyson orbital is practically zero.

As discussed in \cite{Murray2011} (shown also here for completeness), the function $R(\theta_L)$ is given by
	\begin{eqnarray}\label{eq:App15}
		R(\theta_L)&=&\Bigg[F_0(\theta_L)-\frac{4\vert\bar{E}(t_0)\vert}{3\kappa^3}F_2(\theta_L)+\frac{2\vert\bar{E}(t_0)\vert}{3\kappa^3}F_3(\theta_L)\Bigg]^2\nonumber \\ \\
		&&+\frac{2\vert\bar{E}(t_0)\vert}{9\kappa^3}F^2_1(\theta_L)\nonumber,
	\end{eqnarray}
where
	\begin{eqnarray}\label{eq:App16}
		&&F_0(\theta_L)=F(\cos\theta_L,\sin\theta_L), \nonumber \\ \nonumber \\ \nonumber
		&&F_1(\theta_L)=F_v\cos\theta_L-F_u\sin\theta_L, \\ \\ \nonumber
		&&F_2(\theta_L)=F_u\cos\theta_L+F_v\sin\theta_L, \\ \nonumber \\ \nonumber
		&&F_3(\theta_L)=F_{vv}\cos^2\theta_L+F_{uu}\sin^2\theta_L-F_{uv}\sin2\theta_L,
	\end{eqnarray}	
with $F_u$, $F_{v}$, $F_{uu}$, $F_{vv}$, and $F_{uv}$ the first and second order partial derivatives of $F(u,v)$ with
respect to $u$ and $v$, calculated at $u=\cos\theta_L$ and $v=\sin\theta_L$.
The ionization rate we obtain using Eq.~(\ref{eq:App12}) is in very good agreement with the ones obtained in \cite{CDLin2010, Saenz2002}, see Fig.~\ref{fig:IonizationRate}.

\end{document}